\documentclass[twocolumn,letterpaper]{aastex7}

\usepackage{amsmath}

\newcommand{\cs}{c_{\rm s}}

\newcommand{\Rs}{R_{\rm \ast}}
\newcommand{\Ms}{M_{\rm \ast}}

\newcommand{\Mc}{M_{\rm c}}

\begin{document}

\title{Tidal Amplitudes in the Magellanic Cloud Population of Heartbeat Stars}

\author[0000-0002-1417-8024]{Morgan MacLeod}
\affiliation{Institute for Theory \& Computation, Center for Astrophysics, Harvard \& Smithsonian, Cambridge, MA, 02138, USA}
\email{morgan.macleod@cfa.harvard.edu}

\author[0000-0003-4330-287X]{Abraham Loeb}
\affiliation{Institute for Theory \& Computation, Center for Astrophysics, Harvard \& Smithsonian, Cambridge, MA, 02138, USA}
\email{abraham.loeb@cfa.harvard.edu}

\begin{abstract}
Heartbeat (HB) stars exhibit pulses in their light curves once per orbit due to ellipsoidal distortions from strong tides at periapse. We analyze the population of HB stars in the Magellanic Clouds captured by the OGLE survey, and provide broadband spectral energy distribution fitting to estimate physical properties of the HB stars.  The HB stars span a wide range of luminosities, radii, and effective temperatures.  However, we  find that they cluster near loci of strong tidal influence at periapse, indicating that in nearly all cases, strong tides are indeed responsible for their photometric variability.  HB stars tend to populate regions away from the main sequence, where stellar evolution is particularly rapid. We examine the distribution of tidal amplitudes, and show that these can be interpreted through a simplified model of radius growth through stellar evolution and orbital circularization through linear tidal dissipation. When we compare rates of tidal dissipation, we find differences between the modeled modified tidal quality factor among hot ($T_{\rm eff}>6250$~K),  $Q_\ast' \gtrsim 10^7$, and cool ($T_{\rm eff}<6250$~K), $Q_\ast'\sim 10^5$, stars, which is qualitatively consistent with models of efficient tidal dissipation in the convective envelopes of cool stars. We find that this model can reproduce the observed observed locations and amplitudes of HB stars in the Hertzsprung-Russell diagram. The hot stars, in particular, extend to amplitudes near, but not beyond, the threshold for nonlinear tidal wave breaking on stellar surfaces, suggesting a physical saturation of tidal amplitudes at this threshold. 
\end{abstract}


\section{Introduction} \label{sec:intro}

Heartbeat (HB) stars are eccentric binary systems that exhibit distinctive, narrow pulses in their light curves, resembling electrocardiograms \citep{2012ApJ...753...86T}. From theoretical models of these systems, the origin of the variation can be traced to ellipsoidal distortions of the stars each periapse passage  \citep{1995ApJ...449..294K}. Some HB systems also exhibit tidally-excited dynamical oscillations between periapse passages \citep[e.g.][]{2011ApJS..197....4W,2012MNRAS.420.3126F,2012MNRAS.421..983B,2013MNRAS.434..925H,2017ApJ...834...59G,2017MNRAS.472.1538F,2018MNRAS.473.5165H,2019ApJ...885...46G,2019MNRAS.489.4705J,2019ApJ...885...46G,2020ApJ...888...95G,2021A&A...647A..12K,2022A&A...659A..47K,2023A&A...671A..22K,2024MNRAS.530..586L,2024ApJ...974..278L} opening the field of tidal asteroseismology, where tidal perturbations drive a spectrum of nonradial oscillations that can be used to characterize the star's interior structure in detail \citep[e.g.][]{2012MNRAS.421..983B,2014MNRAS.440.3036O,2017MNRAS.467.2494P,2020ApJ...896..161G,2020ApJ...888...95G,2020ApJ...903..122C,2021MNRAS.501..483F,2021FrASS...8...67G,2022MNRAS.517..437G,2023ApJ...945...43S,2024MNRAS.530..586L,2025PASJ...77..118L}.

The discovery of heartbeat stars has been driven by photometric surveys, from systems originally characterized by {\it Kepler} \citep[for an overview, see][]{2011ApJS..197....4W,2012ApJ...753...86T,2023ApJS..266...28L,2024MNRAS.530..586L} to those identified in subsequent space missions like {\it Brite} \citep{2016PASP..128l5001P,2017MNRAS.467.2494P,2019MNRAS.488...64P} and {\it TESS} \citep[e.g.][]{2015JATIS...1a4003R,2019MNRAS.489.4705J,2021A&A...647A..12K,2024ApJ...974..278L,2024MNRAS.534..281L,2025ApJS..276...17S}. Recently, \citet{2022ApJS..259...16W,2022ApJ...928..135W} reported upon and analyzed ground-based photometry from the OGLE survey that lead to the identification of nearly one thousand new HB systems in the galactic bulge and Magellanic clouds. For the subset of the HB systems that have been analyzed in detail, spectroscopic follow up has facilitated orbital modeling to compliment the photometric analysis \citep[e.g.][to give just a few, representative examples]{2011ApJS..197....4W,2016ApJ...829...34S,2021MNRAS.506.4083J}. While spectroscopy is invaluable, some meaningful constraints can be drawn from light curve morphology alone \citep{1995ApJ...449..294K,2012ApJ...753...86T,2022ApJ...928..135W,2023ApJS..266...28L,2024MNRAS.534..281L}. 

By their nature, the typical targets identified by these surveys varied. The {\it Kepler} HB systems focused on A--F stars of roughly solar mass \citep{2012ApJ...753...86T,2023ApJS..266...28L}. The {\it Brite} and {\it TESS} sources highlighted HB behavior among intermediate and massive stars \citep{2017MNRAS.467.2494P,2019MNRAS.488...64P,2019MNRAS.489.4705J,2021MNRAS.506.4083J,2021A&A...647A..12K,2024MNRAS.534..281L}. The OGLE population in the Magellanic clouds favors particularly luminous sources -- either lower-mass giants or massive, hot stars \citep{2022ApJS..259...16W,2022ApJ...928..135W}.  These discoveries have revealed that low-mass HB star amplitudes are typically measured in parts-per-thousand \citep[e.g.][]{2012ApJ...753...86T,2023ApJS..266...28L}, 
while the massive HB systems can have small amplitudes, but also exhibit $\sim 10\%$ variations in some cases \citep{2019MNRAS.489.4705J}, approaching the threshold for nonlinear dissipation \citep{2022ApJ...937...37M,2023NatAs...7.1218M}. 

The amplitudes of tidal oscillations  -- and their photometric counterparts -- are a consequence of the evolutionary processes that lead binary systems to interact. Stellar evolution changes stellar radii, tides dissipate orbital energy \citep[e.g.][]{1981A&A....99..126H,2018MNRAS.476..482V,2021MNRAS.503.5569V}, and multiple-star systems can lead to secular orbital changes \citep[e.g.][]{2007ApJ...669.1298F,2013MNRAS.429.2425F}.
Importantly, the oscillations themselves also couple dynamically to the orbit through mode resonances and dissipation \citep{1999A&A...350..129W,2014MNRAS.443.2957B,2017MNRAS.472L..25F,2017MNRAS.472.1538F,2018MNRAS.473.5165H,2018MNRAS.476..482V,2021MNRAS.501..483F,2021MNRAS.503.5569V,2023Univ....9..514O,2024NatAs...8.1387B}. 

Yet, the relative rates of stellar evolution and dissipation vary widely across stars of different types \citep{2014ARA&A..52..171O}. In general, stellar evolution is more rapid for the  massive stars than it is for low-mass stars. And tidal dissipation is much more effective in turbulent, convective envelopes than in stably-stratified radiative layers \citep[e.g.][]{1977A&A....57..383Z,1981A&A....99..126H,2014ARA&A..52..171O}, it is also strongly frequency-dependent \citep{2014ARA&A..52..171O,2014MNRAS.443.2957B,2017MNRAS.472L..25F,2023ApJ...953...48T,2023ApJ...945...43S}, which is ignored in the weak-friction approximation. Thus, the amplitude of a HB stars tide and its orbital configuration are intimately coupled. 
The emergent trend of higher-HB amplitude massive stars and lower-HB amplitude low-mass stars has led to the theoretical suggestion that massive stars could reach large tidal amplitudes because stellar evolution acts much more rapidly to increase amplitudes than dissipation can decrease them \citep{2020PASA...37...38V,2021MNRAS.503.5569V,2022ApJ...937...37M,2023NatAs...7.1218M}. 
The large and uniform sample of HB stars with known distances in the Magellanic Clouds offers a platform to test these ideas at the interface of tidal excitation and dissipation \citep{2022ApJS..259...16W,2022ApJ...928..135W}. 

In this paper, we draw on the photometric catalog of \citet{2022ApJ...928..135W}, and supplement their results with fitting of the broad-band stellar spectral energy distribution. Combined with a known distance, this lets us convert photometric properties to physical stellar properties, as described in Section 2. In Section 3, we study the distribution of stellar orbits, HB amplitudes, and the emergent distribution of HB sources in the Hertzsprung-Russell (HR) diagram. We show that there are differences between hot and cool stars, that may be traceable to differences in their tidal dissipation. Finally, we examine the nonlinearity limit, and ask whether there are systems approaching nonlinear tidal wave breaking among the sample. In Section 4, we conclude.

\section{Methods}\label{sec:method}
\citet{2022ApJS..259...16W,2022ApJ...928..135W} released and analyzed an unprecedented sample of nearly one thousand HB stars in the Magellanic clouds and galactic bulge. We build on that work by analyzing the 439 LMC and 40 SMC sources. We apply fitting of the stars' spectral energy distributions (SEDs) in order to obtain  estimates of the stellar physical properties like effective temperature, radius, and luminosity.

\subsection{Catalog}
For each star, \citet{2022ApJS..259...16W,2022ApJ...928..135W} catalog the mean magnitude, color, full $V$ and $I$ band photometry, photometric period, along with the fitted parameters of  the \citet{1995ApJ...449..294K} model of flux variations due to the equilibrium tide in an eccentric orbit, 
\begin{equation}\label{k95}
    \frac{\delta F}{F} (t)  = S \times \frac{1 - 3 \sin^2( i ) \sin^2 \left( \varphi(t) + \omega \right) }{ (d(t) / a)^3 } + C,
\end{equation}
where $S$ is an amplitude scale factor, $C$ is a constant offset, $i$ is the orbital inclination, $\omega$ is argument of periastron\footnote{We note the correction of \citet{2022ApJ...928..135W}  to the sign preceding $\omega$ in equation \eqref{k95} relative to earlier works.  }, $a$ is the semi-major axis, and $\varphi(t)$ and $d(t)$ are the time-dependent true anomaly and separation. The variables $\varphi(t)$ and $d(t)$ must be computed by solving Kepler's equation for a given eccentricity, $e$.  The resultant peak-to-peak magnitude variation is cataloged as $A$. The full catalog may be queried in the OGLE collection of variable stars.\footnote{\url{https://ogledb.astrouw.edu.pl/~ogle/OCVS/query.php}} 

For a small subset of stars, we note that the fitted model does not provide a good estimate of the photometric amplitude of the HB signal, and is instead contaminated by eclipses. We use adjusted amplitude estimates for these stars, which include OGLE-LMC-HB-0020, OGLE-LMC-HB-0032, OGLE-LMC-HB-0341, OGLE-LMC-HB-0347, OGLE-LMC-HB-0408, and OGLE-SMC-HB-0007. 

\subsection{SED Fitting}
We supplement the \citet{2022ApJS..259...16W,2022ApJ...928..135W} catalog by performing a fit to the stellar parameters based on its photometric SED. To do so, we make use of {\tt SEDFit v0.6.3} \citep{mkounkel_2023_10436982}, which we slightly modified for the purposes of this work.\footnote{Via a fork of the main code to \url{https://github.com/morganemacleod/SEDFit}} SEDFit queries the UV to IR vizier photometric archive with a cone search for photometry, then matches that photometry with a least-squares fit to a stellar spectral template, varying extinction, $A_V$, distance, $d$, metallicity [Fe/H], $T_{\rm eff}$, and $\log g$. 

We make use of the Kurucz  grid of models provided with SEDFit \citep{1979ApJS...40....1K,1993yCat.6039....0K}. For LMC sources, we fix the distance to $d=49.97$~kpc and the metalicity to [Fe/H] = -0.7. For SMC sources, we apply $d=62.44$~kpc and the metalicity to [Fe/H] = -1.05, where the metalicity estimates are based on the findings of \citep{2021ApJS..252...23S} evaluated at a  distance of $\sim 4$~deg from the LMC or SMC center (see their Figure 11).  To estimate the allowed range of $A_V$ for each source, we query the OGLE-derived reddening map of the LMC and SMC \citep{2021ApJS..252...23S}\footnote{\url{https://ogle.astrouw.edu.pl/cgi-ogle/get_ms_ext.py}} at the location of each HB star to find $E(V-I)$. Though the realistic selective extinction varies, we uniformly adopt $R_V=3.1$ such that $A_V \approx 2.25 E(V-I)$. We set the allowed range between the median reddening minus two sigma and the median reddening plus two sigma from \citet{2021ApJS..252...23S}. 

As a measure against possible conversion differences in photometric systems and telescopes, we implement a minimum uncertainty on each data point's $\nu F_\nu$ (in units of erg~s$^{-1}$~cm$^{-2}$) of $\sigma_e =0.04$~dex or 0.1~mag. This is likely somewhat overestimated, because typical best fit models have $\chi^2/N\lesssim 1$, where $N$ is the number of photometric points. However, we felt this is beneficial because the absolute value of $\chi^2/N$ is not meaningful in our analysis and we prefer that measurement errors represent the possible conversion error in photometric systems rather than statistical errors reported heterogeneously on each observation.

To allow for the automated fitting of all 479 sources, we implement several procedures for handling data irregularities.  For fits that exceed a threshold in $\chi^2 / N>7$, we iteratively reject extreme photometric outliers based on their residual relative to the best-fit model in units of $\sigma_e$, where the residual $R=(M-D)$ and $M$ is the modeled $\log_{10}\nu F_\nu$ and $D$ is the corresponding data. We keep any data $R<5\sigma_e$ from the model. This step is important because outlier photometry can arise either from field crowding or data quality issues, and there are cases were a single outlying point  significantly biases the template fit.   

We additionally check the residual for structure. By default we include a range of 0.2 to 4~microns in wavelength in the SED fit. But in crowed fields or for stars with infrared excess, the data above $\sim 1$~micron sometimes has a correlated slope in the residual. We check for this by evaluating the linear slope of the residual $R$, over the 1 to 5~micron range, measuring the result in dex/micron. Hot stars, with fitted $T_{\rm eff}>6250$~K with slopes larger than 0.1 dex over 4 microns (0.025 dex/micron) are refit including only UV and optical data of wavelength less than 1~micron. We find that cool star estimates of $T_{\rm eff}$ are dramatically improved by including the IR even when some IR excess is present because it locates the peak of the blackbody.  

The result of this fitting procedure is a set of SED-estimated physical stellar properties, which we explore in the following section.  We use the diagonal of the covariance matrix returned from the least-squares fit as a rough estimate of parameter uncertainties. While a case-by-case Monte Carlo fitting would provide more robust posteriors, we report on the median uncertainties briefly as they inform the subsequent interpretation of the derived stellar properties.  The median relative error in effective temperature is $\delta T_{\rm eff}/T_{\rm eff} = 0.06$ (6\%). For the subset of hot ($T_{\rm eff}>6250$~K) stars, it is 13\%. The mean radius error is $\delta \Rs/\Rs =0.03$ (3\%), with a similar factor of two increase for hot stars to 7\%. The median reddening uncertainty, $\delta A_V/A_V = 0.94$, indicating order-unity uncertainty in the estimated reddening (the median value is $\sim0.4$~mag). The median relative uncertainty in $\log g$ is 52\% for cool stars, and it is essentially unconstrained for hot stars.  The radius and effective temperature uncertainties propagate to a median relative luminosity uncertainty of 13\% (26\% for hot stars).

\section{Results}\label{sec:results}

\subsection{Hertzsprung-Russel Diagram}

\begin{figure}[tbp]
    \centering
    \includegraphics[width=\linewidth]{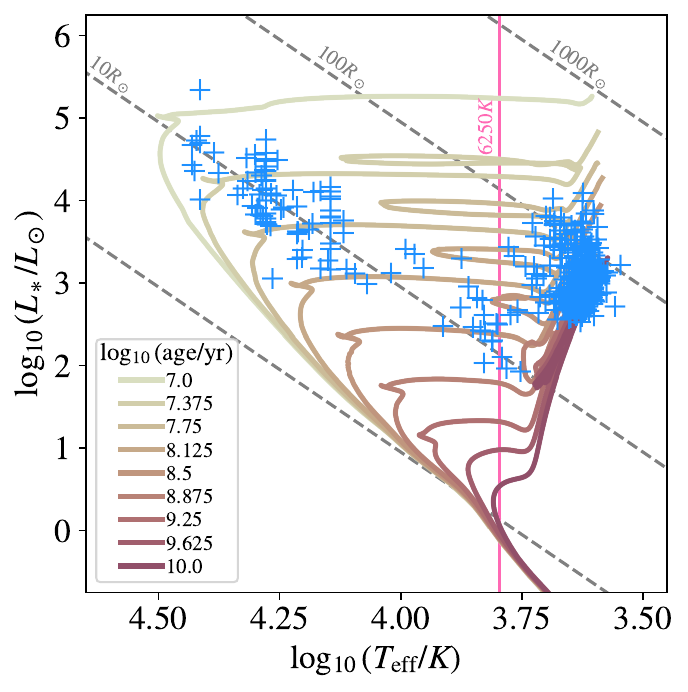}
    \caption{Hertzsprung Russell diagram of the LMC and SMC OGLE population of HB stars, with luminosities and effective temperatures inferred from broad-band SED fitting. Sources are scattered broadly in luminosity and effective temperature, but trace distinct clusters: hot ($\gtrsim 6250$~K) along the horizontal branch and cool ($\lesssim 6250$~K) sources clustered in the giant branch.  }
    \label{fig:HRD}
\end{figure}

Figure \ref{fig:HRD} shows the population of LMC and SMC heartbeat stars cataloged by \citet{2022ApJS..259...16W,2022ApJ...928..135W}. This figure is similar to \citet{2022ApJ...928..135W}'s figure 10 for the LMC sources, where luminosity and effective temperature are estimated from the $UBV$ and $I$ band photometry (see their equations 12 through 16). Our analysis based on fitting the SED of each source to the available photometry is very similar to \citet{2022ApJ...928..135W}'s result, as expected. Typical errors range from smaller than the plotted points for most of the red clump, to somewhat larger than the plotted points for hotter stars, with most of the uncertainty for hot stars arising from reddening versus effective temperature degeneracy. 

Looking at the population, we see that there are HB stars across the HR diagram. Compared to the underlying stellar distribution, which heavily favors low-mass, low-luminosity stars, they are broadly spread in luminosity space, with nearly as many luminous systems as those at lower brightness. The lack of systems below $\sim 10^2 L_\odot$ is because of the survey limiting magnitude. The HB star's appearance is clustered in regions that don't otherwise host the most stars. In particular, there is a cool population in the ``red clump", which is consistent with temperatures of $\sim 4000$~K, and luminosities of $10^3$ to $10^4$~$L_\odot$. These ``Cool" ($T_{\rm eff}<6250$~K) sources as we will classify them in our subsequent analysis, are complimented by a population of hotter stars closer to the main sequence ($T_{\rm eff}>6250$~K, labeled ``Hot" in subsequent figures). However, the hot population does not trace the main sequence, as might be expected based on stellar number density alone. Red or yellow supergiant sources don't seem to appear in the population. The hot sources appear to lie close to the main sequence near $10^5$~$L_\odot$, or for stars in the range of $\sim 30 M_\odot$, but lie substantially yellower, along the horizontal branch (with radii $\sim 10R_\odot$), for lower mass stars at luminosities of $10^3$ to $10^5$~$L_\odot$. 

Recent analysis of data from the Transiting Exoplanet Survey Satellite (TESS) \citep{2015JATIS...1a4003R} has also revealed numerous heartbeat systems \citep[e.g.][]{2021A&A...647A..12K,2024MNRAS.534..281L,2025ApJS..276...17S}. It is worth comparing the properties of these systems briefly to \citet{2022ApJS..259...16W,2022ApJ...928..135W}'s LMC and SMC catalog from OGLE. Figure 6 of \citet{2024MNRAS.534..281L} compares some of these populations to those of HB stars from Kepler \citep[e.g.][and references therein]{2023ApJS..266...28L}. The majority of the TESS and Kepler systems are lower-mass, galactic sources. Nonetheless, they share the trend of enhanced occurrence among bright, hot stars relative to the underlying population. In the HR Diagram, this implies population in the Hertzsprung gap, filling the region $\lesssim 10^2 L_\odot$ that is largely unpopulated for the OLGE LMC and SMC sources due to the survey's limiting magnitude. 

Despite their prevalence off the main sequence in the Hertzsprung gap, at all epochs, HB stars are somewhat rare and are a small fraction of the total population. This is perhaps visually seen most clearly in  \citet{2022ApJ...928..135W}'s figure 9, which shows the HB stars overlain on color-magnitude diagrams of the LMC and SMC. Even at their greatest relative fraction, HB stars are $\lesssim1\%$ of the population of stars of similar brightness and color. This rarity is likely explained by the combination of the likelihood of having a companion at a particular configuration and the lifetime of a HB system once it develops observable tides.

\subsection{Orbital Properties and Tides}

\begin{figure}
    \centering
    \includegraphics[width=\linewidth]{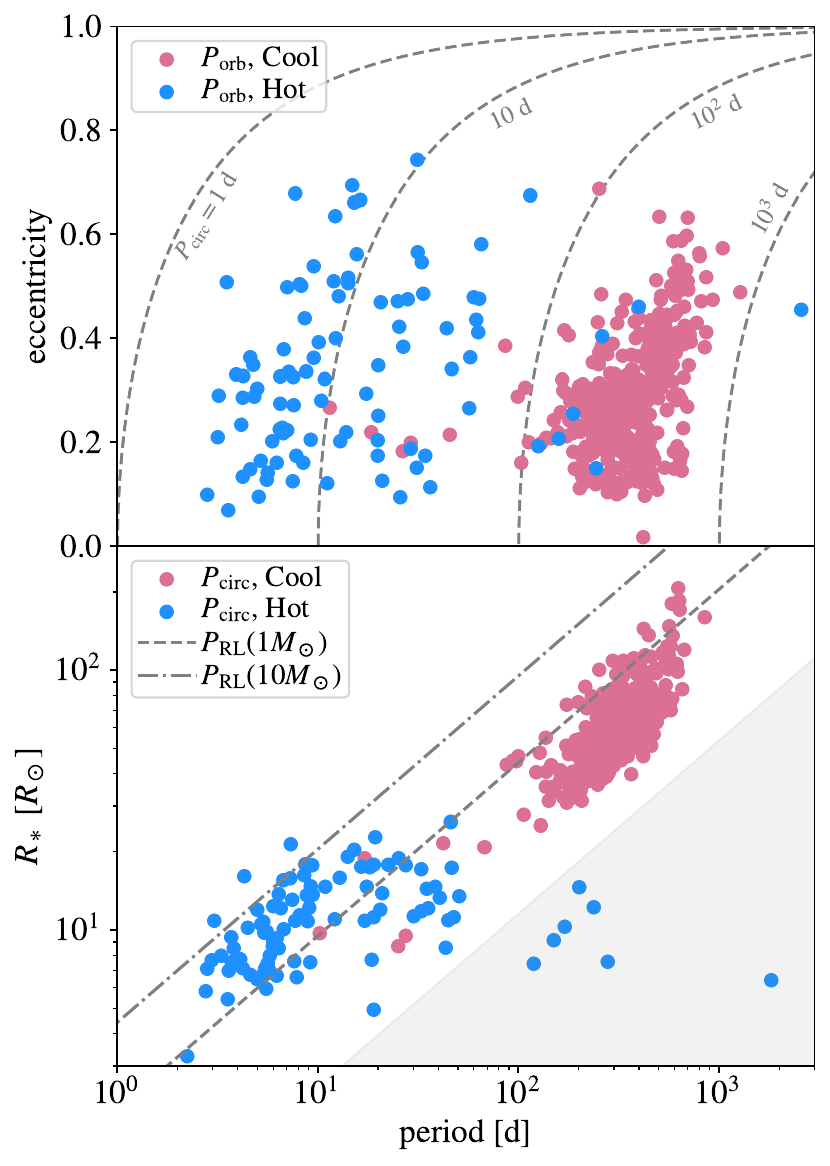}
    \caption{The orbital and tidal properties of the HB population. The upper panel shows modeled orbital eccentricity as a function of period, showing curves of constant angular momentum, which sources might follow as tidal dissipation causes them to circularize. Sources are labeled hot ($T_{\rm eff}>6250$~K) or cool ($T_{\rm eff}<6250$~K) to approximate the location of the Kraft break that divides radiative and convective stellar envelopes.  The lower panel compares sources circularization orbital periods, $P_{\rm circ}$, to stars radii. The dashed lines compare to the Roche limit period, $P_{\rm RL}$ as a function of $\Rs$. Sources cluster near the periods of Roche-lobe filling $P_{\rm circ}\sim P_{\rm RL}$, indicating the importance of tides in the majority of sources. }
    \label{fig:periods}
\end{figure}

With stellar radii estimated from SED analysis (see the contours of Figure \ref{fig:HRD}), we can turn our attention to the scale of HB systems' orbits as compared to the stellar sizes. Figure \ref{fig:periods} analyzes two aspects of the orbital distribution. In the top panel, we show the period--eccentricity distribution, derived directly from \citet{2022ApJ...928..135W}'s model fitting. There is an immediately apparent separation of the hot and cool sources, with the hot sources lying almost uniformly at shorter orbital periods of days to tens of days, while the cooler sources orbit in hundreds of days periods. We overlay this period-eccentricity distribution with lines of constant angular momentum, showing where periods that begin at a certain $(P_{\rm orb}, e)$ combination will circularize if angular momentum is conserved, as is roughly expected via tidal evolution \citep{2024MNRAS.534..281L}. These lines are defined by 
\begin{equation}\label{eq:pcirc}
    e = \sqrt{1-\left(\frac{P}{P_{\rm circ}}\right)^{-2/3}}
\end{equation}
where $P_{\rm circ}$ is the orbital period at $e=0$ \citep[see][figure 5]{2024MNRAS.534..281L}. Figure \ref{fig:periods} shows lines of $P_{\rm circ} = 1$, 10, 100, 1000~d.

In the lower panel, we plot the stellar radius, $\Rs$ as a function of $P_{\rm circ}$, computed by inverting equation \eqref{eq:pcirc}, for each system. Broadly, longer period, cooler stars correspond to larger stellar radii (as can also be inferred from Figure \ref{fig:HRD}). However, we also overlay the period corresponding to the Roche limit -- where the star overflows its Roche lobe, assuming a mass ratio of $q=\Mc/\Ms = 0.3$, where $M_c$ is the mass of the perturbing companion star and $\Ms$ is the mass of the HB star. We estimate the Roche limit separation \cite{1983ApJ...268..368E} as
\begin{equation}\label{aRL}
    \frac{a_{\rm RL}}{\Rs} = \frac{0.6 q^{-2/3} + \ln(1+q^{-1/3})}{0.49 q^{-2/3}}
\end{equation}
and then calculate 
\begin{equation}\label{PRL}
    P_{\rm RL} = 2\pi \left( \frac{a_{\rm RL}^3}{GM}\right)^{1/2}, 
\end{equation}
where $M = \Ms + \Mc$ is the system mass, which is a priori unknown. In Figure \ref{fig:periods}, we show two contours of Roche limit period, one for $M=1M_\odot$, and another for $M=10M_\odot$. In general, the cooler stars cluster relatively close to the $1M_\odot$ contour, consistent with these being older, evolved stars of initial mass from one to a few solar masses. The hotter stars are younger, and more massive, and cluster closer to the higher, $10M_\odot$ Roche limit track. 

These clusters of $P_{\rm circ} \sim P_{\rm RL}$ are highly suggestive that these HB stars are indeed tidally-driven. Were tides not responsible for the photometric variations we observe in these systems, we would not expect the physical correspondence of circularization periods with the tidal Roche limit separation. 

However, Figure \ref{fig:periods} does showcase a few outliers. We highlight a gray region of long periods and small inferred stellar radii that does not appear to be consistent with the tidal explanation of variation. The seven sources in this region are OGLE-LMC-HB-0030, OGLE-LMC-HB-0122, OGLE-LMC-HB-0167, OGLE-LMC-HB-0281, OGLE-LMC-HB-0289, OGLE-SMC-HB-0025, and OGLE-SMC-HB-0033. These sources merit further investigation. It is possible that their radii are significantly underestimated in the current analysis (though this seems unlikely given the magnitude of change needed in some cases). It is also possible that a different mechanism is driving their photometric variability -- for example a wind-wind collision \citep[e.g.][]{2021A&A...650A.147S,2024A&A...686A.199K}.  We exclude these sources from our subsequent analysis leaving a total of 434 LMC and 38 SMC stars.

\subsection{Tidal Amplitudes}

\begin{figure}[tbp]
    \centering
    \includegraphics[width=\linewidth]{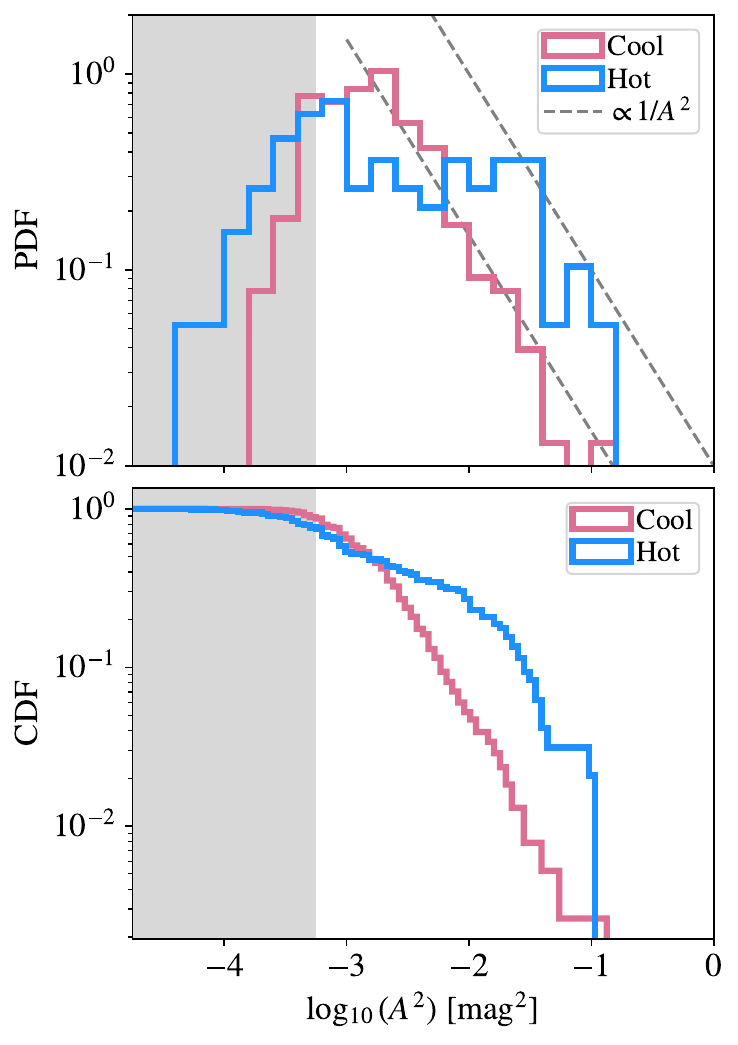}
    \caption{Amplitude distribution of the HB population, divided among cool and hot sources as in Figure \ref{fig:periods}. We count the number of sources in units of $A^2 \propto E_{\rm tide}$. The upper panel shows the probability distribution function (PDF), while the lower panel shows the cumulative distribution function (CDF), counting from large amplitude toward smaller. The gray shaded region marks where sources are likely incomplete due to their small photometric amplitudes. The hot sources spread to higher amplitudes than do the cool, and have a flatter distribution across amplitude. The cool sources are more numerous at low amplitudes and less at higher amplitudes.  Above a certain level in both the hot and cool stars, we see some evidence of a portion of the histogram where the PDF scales as $A^{-2}$, marked by the dashed lines in the upper panel.  }
    \label{fig:Ahist}
\end{figure}

Heartbeat stars gain their once-per-orbit pulses through tidal distortion at periapse. This is ``equilibrium" tide (in the sense of equilibrium with the instantaneous tidal field at periapse) has an order-of-magnitude amplitude 
\begin{equation}\label{DRtide}
    \frac{\delta \Rs}{\Rs} \sim \left(\frac{\Mc}{\Ms} \right) \left(\frac{\Rs}{d} \right)^3,
\end{equation}
where $d$ is the periapse distance in this case. 
 The tidal distortion creates an ellipsoidal stellar shape with polar height $\sim \Rs$, semi-minor axis $\Rs - \delta \Rs$, and semi-major axis $\Rs + \delta \Rs$.  
 To simplest approximation, the flux variation from viewing this ellipsoidal shape at various orientations is proportional to the changing projected surface area, and is therefore $\delta F / F \propto \delta \Rs / \Rs$ \citep[][]{2022MNRAS.516.5021A}. Thus, the photometric amplitude, $A$, measured in the \citet{2022ApJS..259...16W,2022ApJ...928..135W} catalog of HB stars traces, to order of magnitude, the physical amplitude of the tide at peripase. For our purposes, we ignore the possible effects of inclination, and assume 
\begin{equation}\label{Atide}
    A \ {\rm [mag]} \sim \frac{\delta \Rs}{\Rs}
\end{equation}
Thus, the measured photometric amplitude of the tide can be thought of as a rough estimate of the relative tidal amplitude. For example, a photometric amplitude of 0.01~mag implies a roughly 1\% distortion of the star at periapse, while a 0.2~mag amplitude implies a much larger, 20\% physical distortion.  

Like other waves, the energy carried by the tide is proportional to its amplitude squared,  $E_{\rm tide}\propto A^2$. Figure  \ref{fig:Ahist} draws on this proportionality to examine  the amplitude (and tidal energy) distribution of the HB population.  We plot the normalized probability distribution function (PDF) and cumulative distribution function (CDF) of the logarithm of amplitude squared,  $\log_{10}\left( A^2 \right)$, with $A$ in units of $I$-band magnitude. All amplitudes are less than one magnitude, so $\log_{10}\left( A^2 \right) < 0$. 

Figure \ref{fig:Ahist} shows that at low amplitudes, the abundance of sources drops off. We shade this region to denote that portion of the histogram reflects incompleteness given finite photometric precision.  In general hot stars show a broader spread of $A^2$ than do cool stars. At high amplitudes hot stars are similarly abundant at an order of magnitude larger $A^2$  than cool stars, which is easiest to observe in the CDF. Above a certain amplitude, both hot and cool stars appear to follow a roughly $N \propto A^{-2}$ distribution, with offset locations. Cool stars fall onto this curve above $\log_{10}\left( A^2 / {\rm mag}^2 \right) \approx -3$, while hot stars show hints of a similar pattern above $\log_{10}\left( A^2 / {\rm mag}^2 \right) \approx  -1.5$, modulo the lower number statistics of the sources. At smaller $A^2$, the hot star distribution is approximately flat, $N \sim$~constant. The next section considers the possible origin of these amplitude distributions.

\subsection{Shaping of Tidal Amplitudes}

In this section, we sketch a process where the occurrence of heartbeat oscillations comes at the intersection of two processes: stellar evolution driving radius growth, and tidal dissipation driving orbital change. Radius growth is important because it increases the tidal vulnerability, and thus the amplitude of the tide, equation \eqref{DRtide}. Tidal dissipation can also be important because it moderates orbital eccentricity, leading systems toward circular, synchronous configurations, in which they no longer show heartbeat-style pulses due to time-variable tides. It is important to emphasize that the process we discuss here is the simplest possible. For example, orbits can evolve due to secular interaction in hierarchical multiples, not just the tidal dissipation of the pair \citep{2007ApJ...669.1298F}. With that caveat in mind, we can nonetheless establish some baseline expectations from this combination of two processes. 

We imagine main sequence binaries with a broad range of initial eccentricities, initially in configurations in which tides are not important. In a fixed orbit, radius growth is then required for the tidal amplitude to increase. If stellar evolution changes a star's radius on a characteristic timescale $\tau_{\rm evol} = |\Rs/\dot \Rs|$, then over that timescale stellar radii increase, and on a similar timescale, and so does the tidal amplitude, $A \propto (\Rs/d)^3$. The timescale for amplitude increase in a fixed orbit is $\tau_{\rm evol,A} = |A/\dot A| = \tau_{\rm evol}/3$. 

However, the tidal amplitude cannot grow arbitrarily large without reducing the eccentricity of the orbit. 
We have discussed the energy required to raise the tidal distortion, $E_{\rm tide} \propto (\delta \Rs / \Rs)^2 \propto A^2$. 
This energy is derived from the orbital energy, and, in a perfectly conservative system, this energy is returned as the tidal distortion relaxes after a periapse passage. However, if a fraction of the energy of the distortion is dissipated each orbit, then orbital energy is lost and the orbit begins to circularize. A linear dissipation  is described by a characteristic dissipation timescale, $\tau_{\rm diss} = |E_{\rm tide}/\dot E_{\rm tide}|$, rather than the tidal amplitude \citep[e.g.][]{1977A&A....57..383Z,1981A&A....99..126H,1989A&A...220..112Z}. The orbit will change on a related timescale, $\tau_{\rm circ} = |E_{\rm orb}/\dot E_{\rm orb}|$, where we associate $\dot E_{\rm orb}$ and $\dot E_{\rm tide}$.  This means that the orbital evolution rate is proportional to the tidal energy, and thus to the amplitude squared, $|\dot E_{\rm orb}| = |\dot E_{\rm tide}| = E_{\rm tide} / \tau_{\rm diss} \propto A^2 / \tau_{\rm diss}$ \citep[see][for a related example]{2025ApJ...981...77M}.  

In particular, the combination of stellar evolution and tidal dissipation imply that there will be two portions of the amplitude distribution: 
\begin{enumerate}
    \item  In one case, the amplitudes are shaped by stellar evolution, where $\tau_{\rm evol}\ll \tau_{\rm circ}$. Here, orbital change due to tidal dissipation is slow relative to stellar evolution. In this regime tides can continue to grow in amplitude as the star evolves. The abundance of systems depends on the evolution rate but not the amplitude, with $N \propto \tau_{\rm evol} \propto A^0$. 
    \item In the other case, tidal dissipation shapes the amplitude distribution, where $\tau_{\rm evol}\gg \tau_{\rm circ}$. When tidal dissipation becomes the fastest evolutionary driver of the system, the orbital eccentricity decreases at roughly fixed amplitude, eventually removing a system from the heartbeat population. In this regime, there is a steady-state population of binary systems with evolving orbits. The system lifetime is inversely proportional to its evolution rate, and thus the abundance of systems at a given amplitude is $N\propto \tau_{\rm circ} \propto A^{-2}$. 
\end{enumerate} 
The division between these regimes comes from equating the rate of amplitude growth, $\tau_{\rm evol}/3$, with the rate of orbital evolution, $\tau_{\rm circ}$. We find $A_{\rm diss}^2 \propto \tau_{\rm diss}/\tau_{\rm evol}$, where $A_{\rm diss}$ is the amplitude at which tidal dissipation becomes important. Slower stellar evolution, or more-rapid dissipation both limit the population to smaller critical amplitudes. Thus, all else being equal, case (1) above comes at small $A$, while case (2) comes at large $A$.

We can re-examine Figure \ref{fig:Ahist} in light of this theory. We associate the declining abundance of systems at high amplitudes to shaping of the population by tidal dissipation (case 2). We attribute the approximately flat population at lower amplitudes to amplitude increase due to stellar evolution (case 1). We interpret this feature as evidence for tidal shaping of the HB star population, especially among the cool sources. The cool stars show an extended tail spanning roughly two orders of magnitude in tidal energy, where $N \propto A^{-2}$, consistent with lifetime expectations if these systems are undergoing tidal circularization at rates dependent upon their amplitudes. By contrast, only the highest amplitude hot stars show similar evidence, with the rest of the distribution consistent with being undisturbed by tidal evolution. In the following section, we expand our discussion to quantitatively compare dissipation between hot and cool HB stars.

\begin{figure}[tbp]
    \centering
    \includegraphics[width=0.95\linewidth]{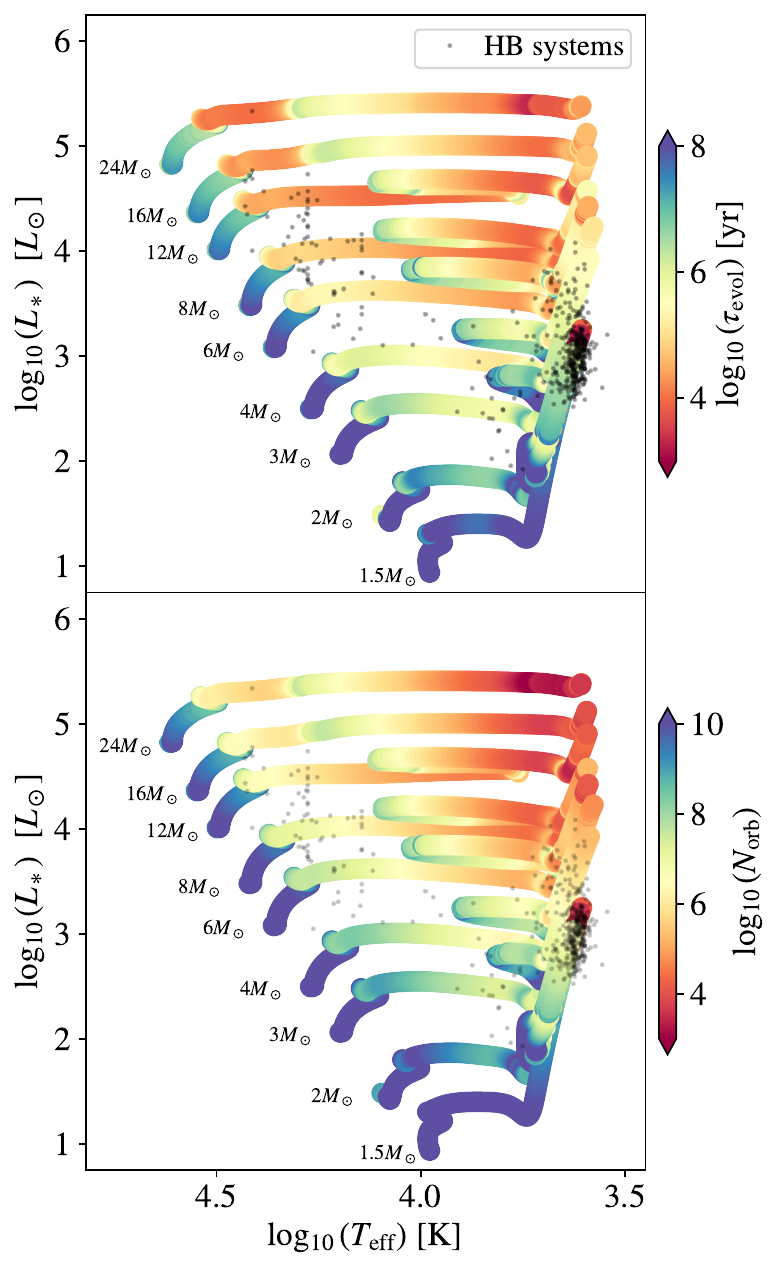}
    \caption{Stellar evolution timescales and comparison to binary orbital periods. The upper panel shows $\tau_{\rm evol}=|\Rs/\dot \Rs|$, the characteristic timescale for the radius to change. The lower panel divides by the circular orbital period at the Roche limit, such that $N_{\rm orb}=\tau_{\rm evol}/P_{\rm RL}$.  In the upper panel, we see that stars evolve slowly near the main sequence, exhibiting long $\tau_{\rm evol}$. As they cross the horizontal branch, they evolve more rapidly, and $\tau_{\rm evol}$ decreases. In general, with their shorter lifetimes, massive stars evolve more rapidly than lower mass stars. The $N_{\rm orb}$ panel shows similar trends, but emphasizes that in some configurations (like near the main sequence) stars might go through $10^{10}$ orbital cycles before stellar evolution leads to a change in stellar radius and tidal amplitude, while for evolved stars, the lifetime in a given configuration can be as short as $\sim 10^4$ orbits. The overlaid HB systems are not generally found where $\tau_{\rm evol}$ or $N_{\rm orb}$ are large.  }
    \label{fig:HRD_evol}
\end{figure}

\subsection{Linear Tidal Dissipation and Quality Factor}

We can trace the combination of stellar growth and tidal dissipation further to understand the occurrence of HB systems in the HR diagram, as described in the previous section. 

To assess the relative timescales available for stellar evolution to increase HB amplitudes, $\tau_{\rm evol}$, in Figure \ref{fig:HRD_evol}, we turn to theoretical stellar tracks and overplot the observed HB population. The upper panel of Figure \ref{fig:HRD_evol} shows $\tau_{\rm evol} = |\Rs/\dot \Rs|$, along stellar tracks of various masses. The lower panel shows $N_{\rm orb}= \tau_{\rm evol}/P_{\rm RL}$, the approximate number of orbits completed over the timescale of $\tau_{\rm evol}$, taking the $e\rightarrow 0$ limit and $a=a_{\rm RL}$, equation \eqref{aRL}. Stars evolve slowest on the main sequence, changing much more rapidly in size as they evolve, cross the horizontal branch, and become giants. In spaces coincident with the population of HB stars, we see $\tau_{\rm evol}\sim 10^5$ to $10^6$~yr.  This corresponds to as many as $10^{10}$ orbits near the main sequence, or as few as $10^3$ orbits for red supergiants.

A priori, the tidal dissipation timescale is more uncertain. One of the simple parametrization is through a tidal quality factor $Q_\ast$, that relates the energy stored in the tide to the cumulative dissipation over one radian of tidal cycle \citep{1966Icar....5..375G,1999ssd..book.....M}. The modified tidal quality factor is $Q_\ast'= 3Q_\ast/2k_2$, where $k_2$ is the potential Love number of the quadrupole mode that corresponds to the tidal field, and accounts for the centrally concentrated nature of stellar envelopes \citep[e.g.][]{1909RSPSA..82...73L,2007ApJ...661.1180O,2014ARA&A..52..171O};  it is proportional to the squared overlap integral between the tidal potential and stellar density profile  \citep{2020MNRAS.496.3767V}.  Typical values estimated from the circularization of binary star systems are $Q_\ast'\sim 10^6$, but Hot Jupiters around sun-like stars lead to estimates of $Q_\ast'\sim10^8$ at short periods \citep[e.g.][]{2014ARA&A..52..171O,2018AJ....155..165P,2020MNRAS.498.2270B,2025ApJ...981...77M,2025arXiv250113929D}. This is an important reminder that this simple parametrization of $Q_\ast'$ almost certainly depends on stellar type, dissipation mechanism, and forcing frequency \citep[for example, see the synopsis of][for sun-like stars]{2020MNRAS.498.2270B}. 

Given a value of $Q_\ast$, we estimate the dissipation timescale $\tau_{\rm diss}= |E_{\rm tide}/\dot E_{\rm tide}| = Q_\ast P_{\rm orb}/2\pi$, where we have associated the tidal period with the forcing period of the orbit for the equilibrium tidal distortion \citep{}. With $\tau_{\rm diss}$, we can now estimate $\tau_{\rm circ}$ given $E_{\rm tide}$. Quantitatively,
\begin{equation}
    \tau_{\rm circ} = \left|\frac{E_{\rm orb}}{\dot E_{\rm tide}}\right| = \left|\frac{E_{\rm orb}}{ E_{\rm tide}}\right| \tau_{\rm diss}
\end{equation}
The orbital energy is $E_{\rm orb}=G M_*^2 q/2a$. The tidal energy depends on the stellar binding energy and stellar structure. To a very crude approximation, we express $E_{\rm tide}$ in terms of the photometric amplitude as $E_{\rm tide} \sim (\delta R_*/R_*)^2 k_2 GM_*^2/R_* \sim (A/{\rm mag})^2 k_2 GM_*^2/R_*$. We then have \citep[see equations 5--10 of][]{2014ARA&A..52..171O},
\begin{equation}
    \tau_{\rm circ} \sim q \frac{\Rs}{a} \frac{Q_\ast' }{A^2} \frac{P_{\rm orb}}{2\pi},
\end{equation}
where $A$ is expressed in magnitudes.

To solve for the amplitude at which dissipation becomes important, we equate the orbital circularization and stellar evolution timescales,  $\tau_{\rm circ} = \tau_A = \tau_{\rm evol}/3$. Solving for $A$, we find
\begin{equation}\label{Adiss}
A_{\rm diss} \sim \sqrt{\frac{3q \Rs}{2\pi a} \frac{Q_\ast' P_{\rm orb}}{\tau_{\rm evol}}} \sim    0.3\sqrt{ \frac{Q_\ast' P_{\rm orb}}{\tau_{\rm evol}}}, 
\end{equation}
where the second similarity adopts representative values of  $q\sim 1/3$ and $\Rs/a\sim 1/2$.  We note that this more quantitative derivation still leads to the scaling mentioned in the previous section $A_{\rm diss}^2 \propto \tau_{\rm diss}/\tau_{\rm evol}$.  

Given a known value of $A_{\rm diss}$, we can invert equation \eqref{Adiss} and solve for $Q_\ast'$,
\begin{equation}
    Q_\ast' \sim \frac{2 \pi}{3q} \frac{a}{\Rs} \frac{\tau_{\rm evol}}{P_{\rm orb}} A_{\rm diss}^2
\end{equation}
For our typical hot star population, $A_{\rm diss}^2 \gtrsim 0.1$~mag${^2}$, because Figure \ref{fig:Ahist} shows little or no evidence of a decay-limited population above this threshold. Plugging in  representative values of $q=1/3$, $\Rs = 10 R_\odot$, $\Ms = 10 M_\odot$, we see $\tau_{\rm evol}\sim10^5$~yr. For this combination, $P_{\rm RL}\sim10^{-2}$~yr, we find $Q'_{\rm hot} \gtrsim 3\times 10^7$. For our cool star population, we might use representative values of $\Ms= 2M_\odot$, $\Rs = 100 R_\odot$, giving $P_{\rm RL}\sim 1$~yr and $\tau_{\rm evol}\sim 10^6$~yr. For the cool population, Figure \ref{fig:Ahist} shows $A_{\rm diss}^2 \sim 10^{-2.5}$~mag${^2}$. Solving for $Q_\ast'$, we find $Q'_{\rm cool}\sim 10^5$. Thus, the modified tidal quality factor in these two populations is different by several orders of magnitude.

The difference in $Q_\ast'$  points to weaker dissipation  among the hot stars. This sort of distinction tracks with differences in stellar structure and the implied dissipation mode. The cool stars, below the Kraft break, will have massive convective envelopes and efficient tidal dissipation as the ordered field of the tide interacts with the disordered motions of convection. The hot stars tend to have radiative envelopes and dissipation either occurs in their convective cores or through radiative damping near the photosphere, either of which is expected to be much less efficient than convection \citep{1977A&A....57..383Z,2014ARA&A..52..171O,2020MNRAS.498.2270B}. 

Theoretically, the effective tidal quality factor of the convective envelopes can be estimated based on the convective timescale \citep{1992RSPTA.341...39P,1995A&A...296..709V,2020MNRAS.496.3767V,2021MNRAS.503.5569V}, 
\begin{equation}
    \tau_{\rm conv}\sim \left( \frac{\Ms \Rs^2}{L_\ast}\right)^{1/3},
\end{equation}
which implies a dissipation timescale
\begin{equation}
    \tau_{\rm diss} \sim \tau_{\rm conv} \left(\frac{\omega_{\ast}}{\Omega_{\rm orb}}\right)^2,
\end{equation}
where $\omega_\ast^2 = G\Ms/\Rs^3$ is the approximate free-oscillation frequency of the star in response to the tidal distortion. The tidal quality factor is then
$Q_{\rm mlt}\sim \tau_{\rm diss} \Omega_{\rm orb} \sim \tau_{\rm conv} \omega_{\ast}^2 /\Omega_{\rm orb}$ \citep[e.g. Section 4.9 of][]{1999ssd..book.....M}, or 
\begin{equation}
    Q_{\rm mlt}' \sim \frac{3}{2k_2}  \frac{\omega_{\ast}^2}{ \Omega_{\rm orb}} \tau_{\rm conv}.
\end{equation}
where the subscript refers to the mixing-length theory of convection \citep{2014ARA&A..52..171O}. However, when the convective overturn time is long compared to the forcing period of the tide, $t_{\rm conv} > \Omega_{\rm orb}^{-1}$, the effective dissipation is reduced and $Q_\ast'$ increases by a factor of roughly
\begin{equation}\label{GN}
    f= 1 + \left(\frac{\omega_{\rm tide}}{\omega_{\rm conv}} \right)^2
\end{equation}
where $\omega_{\rm tide}$ is the tidal frequency, which we have been approximating with $\Omega_{\rm orb}$, and $\omega_{\rm conv} = 1/\tau_{\rm conv}$ is the convective overturn frequency \citep{1977Icar...30..301G}. While  equation \eqref{GN} captures the limiting behavoir in high and low-frequency regimes, more recent numerical work has better captured intermediate behavior \citep{2020MNRAS.497.3400D} and offers a different fit to $f$ based on simulation results. 
The effective quality factor can then be estimated as
\begin{equation}
    Q'_{\ast} \sim f Q'_{\rm mlt}.
\end{equation}

We can now use these derivations to estimate representative quality factors and compare them to our derivations from the HB systems. 
 For $L_\ast \sim 10^3 L_\odot$, $\Ms= 2M_\odot$, $\Rs = 100 R_\odot$, $k_2 \sim 0.03$, we estimate $Q_\ast' \sim 7\times10^5$ (or $4\times10^4$ with the fitted $f$ of \citet{2020MNRAS.497.3400D}, the inverse of their equation 13). Thus, the empirical derivation of $Q_{\rm cool}'\sim 10^5$ for the cool stars is consistent with the theoretical predictions for convectively turbulent dissipation. However, convective envelopes aren't generally present at $T_{\rm eff}\gtrsim6250$~K, and the radiative envelopes of hot stars have predicted $Q_\ast'\gg 10^8$ due to radiative damping of g-modes \citep[e.g.][]{1975A&A....41..329Z,2007ApJ...661.1180O,2020MNRAS.498.2270B,2021MNRAS.503.5569V,2023MNRAS.524.5575P}. In practice, effective dissipation seems to be much stronger, with populations of binaries (of, admittedly, lower mass) supporting as low as $Q_\ast'\sim 10^6$ \citep{2023MNRAS.524.5575P}. The hot star population derivation of $Q'_{\rm hot}\gtrsim3\times 10^7$ is in the broadly-expected parameter range, but is not tightly predicted theoretically -- in part because large star-to-star differences are possible \citep{2014ARA&A..52..171O}.  Therefore, while studies of tides in individual HB systems have been very fruitful to date, the population of HB stars provides a novel means of probing the strength of linear tidal dissipation across a wide range of stellar types \citep[e.g. as also discussed by][]{2025arXiv250113929D}.

\subsection{Occurrence in The HR Diagram}

\begin{figure*}[tbp]
    \includegraphics[width=\textwidth]{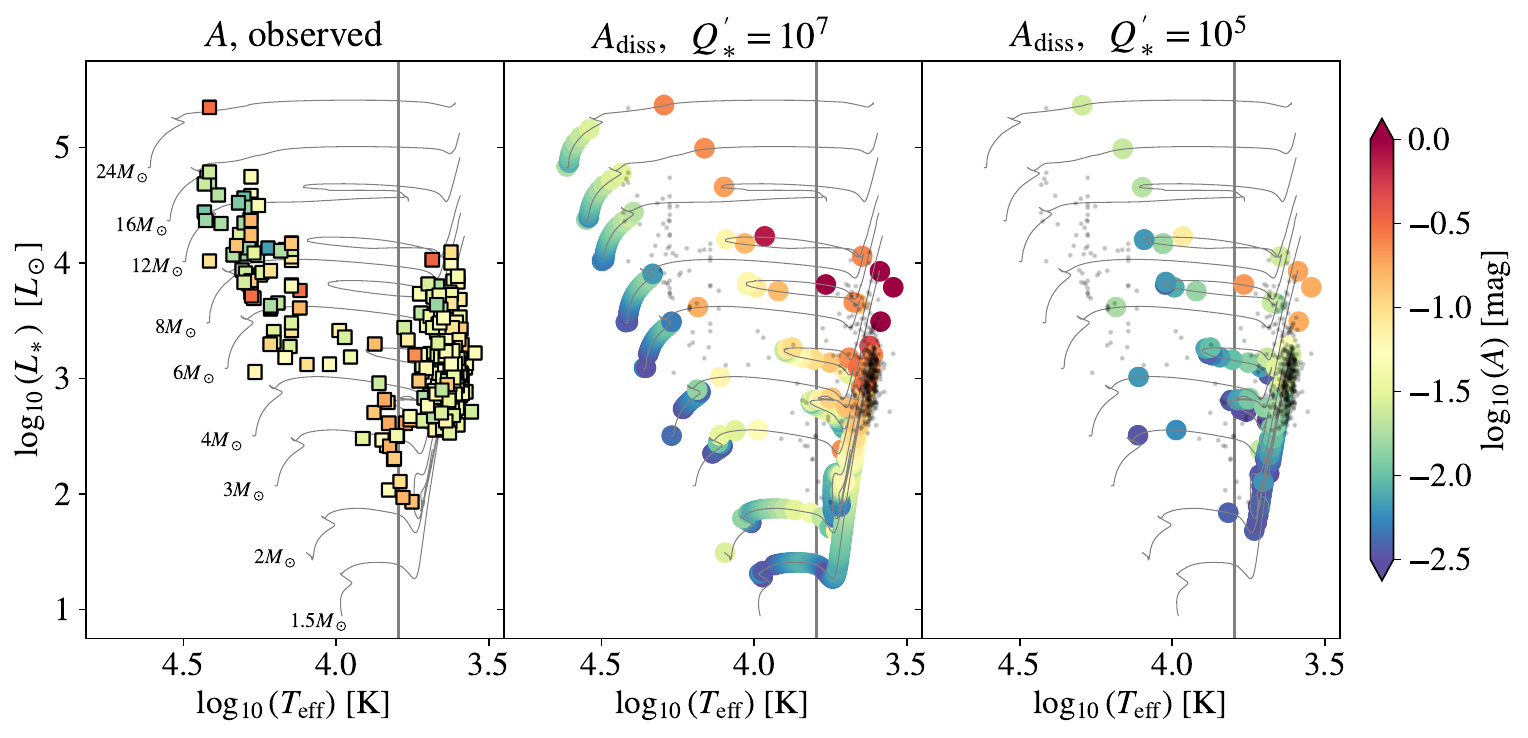}
    \caption{The occurrence and amplitudes of HB stars in the HR diagram. The left panel shows the observed amplitudes of the HB stars, with think lines showing evolutionary tracks, and a vertical line marking $T_{\rm eff}=6250$~K. The middle panel computes the dissipation amplitude $A_{\rm diss}$, for $Q_\ast'= 10^7$, equation \eqref{Adiss}. The right panel adopts   $Q_\ast'= 10^5$. The center and right panels sample points from along the stellar evolutionary tracks at even, 1Myr intervals. This sampling reproduces the distribution and approximate amplitudes of the HB sources. We see HB systems among the hot stars are reproduced well by $Q_\ast' \gtrsim 10^7$. In particular, the combination of the evolution rate and predicted $A_{\rm diss}$ reproduces the tilt of the HB sources relative to the main sequence. post-MS massive stars are comparatively rare  because of their short lifetimes. But stars nearer to the MS do not appear as HB systems because $A_{\rm diss}$ is too small to be observable. The right panel, with $Q_\ast'=10^5$ better predicts the amplitudes of the low-mass giants, but is incompatible with all of the hot HB systems, predicting $A_{\rm diss}<10^{-2.5}$~mag.   }
    \label{fig:HRD_pop}
\end{figure*}

Having discussed the how stellar evolution and tides might shape a population of HB star systems, we now turn our attention to the distribution of that population in the HR diagram (Figure \ref{fig:HRD}). In Figure  \ref{fig:HRD_pop}, we show $A_{\rm diss}$ for an adopted values of $Q_*'= 10^5$ and $Q_*'= 10^7$. To do so, we approximate $P_{\rm orb} \sim P_{\rm RL}$, the circular orbit limit, given in equation \eqref{PRL}. The amplitudes of the HB systems are shown in the colorscale, as are $A_{\rm diss}$ for the model stellar tracks of various mass. The model tracks are sampled at even, $1$~Myr intervals, as a crude representation of the relative abundance of stars at different locations in the HR diagram. We note that this sampling represents the fact that regions of the HR diagram where that a star evolves quickly through will be less populated, but we do not weight according to the fact that lower-mass stars are more common than high-mass stars. 

Examining Figure \ref{fig:HRD_pop}, we see that where $\tau_{\rm evol}$ is long (Figure \ref{fig:HRD_evol}), $A_{\rm diss}$ becomes small, as seen in equation \eqref{Adiss}. Thus, along the main sequence, where stars spend a long time without major radius change, the characteristic dissipation amplitude is very low. Similarly, for lower-mass stars $A_{\rm diss}$ is lower than for higher-mass stars because their evolution rate is slower. This means that in these slow evolving systems, before the tidal amplitude can reach very high values, dissipation acts to circularize the binaries orbits and remove them from the HB population. What we find, however, is that $A_{\rm diss}$ reaches larger values in massive stars, and off of the main sequence in the horizontal branch. 

For a representative $A_{\rm diss}\sim 0.03$~mag, this threshold is reached nearer the main sequence among massive stars (at higher $T_{\rm eff}$), and  further from the main sequence in lower-mass stars (at lower $T_{\rm eff}$). We can see in Figure \ref{fig:HRD_pop} that this behavior reproduces the ``tilt" of the HB population relative to the main sequence in the HR diagram, originally noted by \citet{2022ApJ...928..135W}. Figure \ref{fig:HRD_pop} shows us that the occurrence of HB stars in the comparatively unpopulated horizontal branch is not coincidental, and in fact relates deeply to their physical mechanism. HB stars favor locations just off the main sequence among the most massive stars, these conditions yielding sufficiently rapid stellar evolution.

Stars also evolve rapidly on the giant branch, producing large $A_{\rm diss}$ even for lower-mass giants. We note that the red supergiants are largely missing from the population not because they are unable to yield large tidal amplitudes, but because their short lifetimes make them rare. Represented in terms of the fixed sampling intervals of  Figure \ref{fig:HRD_pop}, these stars not present in the model sample because they evolve too rapidly through that portion of the HR diagram. However, in a different, selectively-focused sample, one might in principle be able to discover HB stars even among RSGs.

These results show that amplitudes of the LMC and SMC HB population are broadly shaped by stellar radius growth driven by stellar evolution and damping by tidal dissipation. In the parameter space where tidal dissipation is slow relative to stellar evolution, HB stars of particularly large amplitude can be generated.

\subsection{Nonlinear Tidal Dissipation: Tidal Wave Breaking}

\begin{figure*}[tbp]
    \centering
    \includegraphics[width=0.85\linewidth]{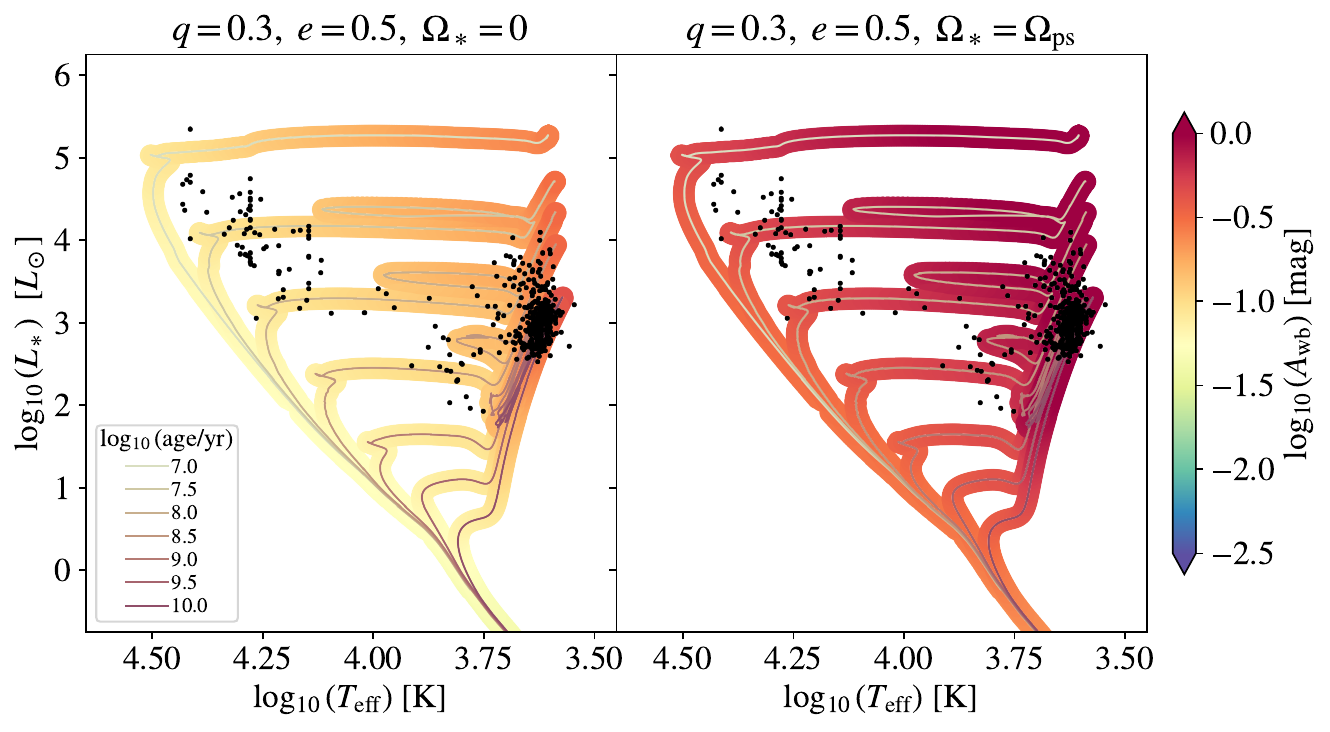}
    \caption{Critical amplitude for nonlinear tidal wave breaking on the stellar surface. Each panel plots $A_{\rm wb}$, equation \eqref{Acrit}, under representative assumptions. The left panel assumes non-spinning HB stars, while the right panel assumes pseudosynchronous rotation. In general we see that $A_{\rm wb}$ is smallest near the main sequence (among the hot stars), and largest on the giant branch (among the cool stars). Overlaid points show the HB systems. A comparison to Figure \ref{fig:HRD_pop} shows that $A_{\rm wb}$ is generally much larger than $A_{\rm diss}$, implying that for most HB stars linear tidal dissipation is important before systems reach nonlinear amplitudes. The exceptions to this statement come for asynchronously rotating, hotter stars.   }
    \label{fig:AcritHRD}
\end{figure*}

\begin{figure}[tbp]
    \centering
    \includegraphics[width=\linewidth]{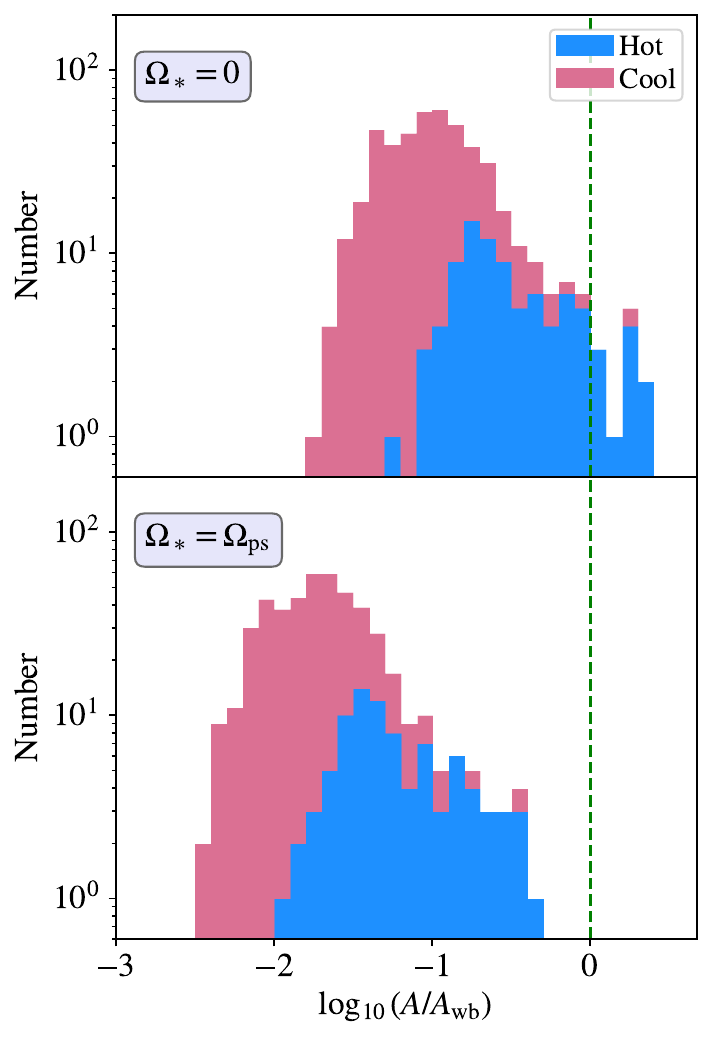}
    \caption{Number of HB systems, stacked for hot and cool sources, as a function of fraction of the wave breaking amplitude. Hot stars generally extend closer to $A_{\rm wb}$. The upper panel assumes nonspinning stars while the lower panel adopts pseudosynchronous rotation. When stars are assumed to be synchronously rotating, $A_{\rm wb}$ appears to be a true physical saturation level in that HB systems approach, but do not cross that threshold.  }
    \label{fig:AcritHist}
\end{figure}

When tidal dissipation does not act to circularize HB systems at small tidal amplitudes, the possibility exists for the time-dependent tidal forcing to become rapid enough that wave breaking occurs \citep{2022ApJ...937...37M,2023NatAs...7.1218M}. This happens when the oscillatory motion of the tide is faster than the sound speed. We consider breaking of the tide forced each periapse passage, which has largest amplitude near the stellar surface, which is also where the sound speed is lowest \citep{2022ApJ...937...37M}. 

In the fluid-frame, the radial speed of the tide is 
\begin{equation}
    v_{\rm tide} \approx \delta \Rs \left( \Omega_{\rm peri} - \Omega_\ast \right),
\end{equation}
where $\delta \Rs$ is the amplitude and
\begin{equation}
    \Omega_{\rm peri} =  \frac{ (1+e)^{1/2}}{ (1-e)^{3/2}} \Omega_{\rm orb},
\end{equation}
$\Omega_{\rm orb}=\sqrt{GM/a^3}$, and $\Omega_\ast$ is the stellar spin rate. A priori the spin of the star is unknown. We can approximate the range of possibilities by considering non-spinning stars, $\Omega_\ast = 0$, or stars that spin at the pseudo-synchronous rate 
\begin{equation}
    \Omega_{\rm ps} = \frac{1 +  \frac{15}{2} e^2 + \frac{45}{8}e^4 + \frac{5}{16}e^6 }{ (1+3e^2 + \frac{3}{8} e^4) (1-e^2)^{3/2} } \Omega_{\rm orb}
\end{equation}
which would result from frequency-independent tidal dissipation through the eccentric orbit \citep{1981A&A....99..126H,2022ApJ...933...25P}. In practice, because a stars' moment of inertia is typically small compared to that of the orbit, orbits can be (pseudo)synchronous without being circular. 

Tidal waves break, cascading into shocks and turbulence when motion is supersonic, 
$v_{\rm tide} > c_s$,  where $c_s$ is the local gas sound speed \citep{2022ApJ...937...37M}. At the surface, we can estimate
\begin{equation}
    c_s \approx \left( \frac{k T_{\rm eff}}{\mu m_p} \right)^{1/2} 
\end{equation}
where $k$ is Boltzmann's constant, $\mu\approx 0.61$ is the mean molecular weight, and $m_p$ is the proton mass. In terms of the photometric amplitude, $v_{\rm tide}\sim A \Rs \left( \Omega_{\rm peri} - \Omega_\ast \right)$ because $\delta \Rs/\Rs \sim A/{\rm mag}$. Thus a critical amplitude for wave breaking is 
\begin{equation}\label{Acrit}
   \left[ \frac{\delta \Rs}{\Rs}\right]_{\rm wb} \sim A_{\rm wb} \sim \frac{c_s}{\Rs \left( \Omega_{\rm peri} - \Omega_\ast \right)},  
\end{equation}
which can be computed for a given orbit and star system. 

Figure \ref{fig:AcritHRD} estimates $A_{\rm wb}$ for theoretical stellar isochrones, assuming in each case an eccentric orbit with $q = 0.3$ and $e=0.5$. The two panels adopt $\Omega_\ast=0$ and $\Omega_\ast = \Omega_{\rm ps}$, respectively. In each case we overplot the HB population for reference.  Comparing the two panels, we see that $A_{\rm wb}$ is always larger in pseudosynchronous rotation than it is for non rotating stars. This is because the pseudosynchronous rate is not equal to but approaches $\Omega_{\rm peri}$, reducing the apparent tidal frequency from the perspective of the stellar fluid. Secondly, giant branch stars have higher $A_{\rm wb}$ than do their main sequence counterparts. This is because at constant luminosity, $\cs \propto \Rs^{-1/4}$, while the tidal velocity at fixed relative amplitude scales with $\Rs^{-1/2}$. Thus, $A_{\rm wb}\propto\Rs^{1/4}$ at fixed luminosity. 

For most of the HB stars seen, $A \ll A_{\rm wb}$. However, \citet{2023NatAs...7.1218M} have argued that breaking waves mark each periapse passage of OGLE-LMC-HB-0254, also known as MACHO 80.7443.1718 \citep[though an alternate perspective is offered by][]{2024A&A...686A.199K}. Figure \ref{fig:AcritHist} examines the distribution of HB stars in $A/A_{\rm wb}$, adopting the modeled orbital eccentricities, stellar radii, and photometric amplitudes. We see that $A/A_{\rm wb}$ is uniformly smaller for the pseudosynchronous rotation case (lower panel), which we also suggest is the more realistic option given that some degree of spin up by tidal dissipation is likely, even if not at to the particular pseudosynchronous rate. 

We also note that cool stars are generally restricted to $A/A_{\rm wb}\lesssim0.1$ for $\Omega_\ast = \Omega_{\rm ps}$, while hot stars extend to $A \sim A_{\rm wb}$. This is likely another reflection of the tidal sculpting of the HB population described previously -- it appears that $A_{\rm diss} \ll A_{\rm wb}$ for cool stars, which means that  tidal dissipation is effective at removing cool, eccentric binaries from the HB population before their amplitudes grow above $\sim 0.1 A_{\rm wb}$. 

Finally, Figure \ref{fig:AcritHist} emphasizes that $A_{\rm wb}$ (with $\Omega_\ast = \Omega_{\rm ps}$) likely is a true physical limit. We see a decreased abundance of systems approaching this limit and no systems beyond it. This is as expected if wave breaking saturates wave amplitudes to $A\lesssim A_{\rm wb}$, effectively providing very high dissipation rates once the wave breaking threshold is reached \citep{2022ApJ...937...37M}. Further investigation of individual sources nearest the threshold will reveal if there are multiple sources in the HB population with actively breaking waves, constraining dissipation during this nonlinear phase.

\section{Conclusions}\label{sec:conclusions}

The \citet{2022ApJS..259...16W,2022ApJ...928..135W} sample of 1000 heartbeat stars in the Magellanic clouds and galactic bulge is a landmark not just in number of sources, but in the prevalence of massive stars and stars across the color magnitude diagram. With their known distances, the LMC and SMC sources  are particularly suited to physical characterization and are the focus of our study. We perform SED fits based on the available photometry in order to discuss the physical origin and amplitudes of HB variations among these systems. Some key findings of this paper are:
\begin{enumerate}
    \item The population of HB stars in the LMC \& SMC show a range of stellar types including hot, main sequence or horizontal branch stars and cool giants, appearing in locations in the HRD where stellar evolution is particularly rapid compared to tidal dissipation.
    \item The vast majority of systems lie in modeled orbits that place their periapse distances close enough to lead to strong tidal distortions, validating the eccentric-orbit plus tides physical origin of their variability.  
    \item Both hot and cool stars show a spread in amplitudes as expected from increasing tidal strength due to stellar evolutionary radius growth in eccentric pairs. Hot stars however, extend to higher amplitudes, while cool stars show evidence of a power-law abundance $N \propto 1/A^2$, evidence that the lifetime of these heartbeat systems is inversely proportional to their tidal energy because tidal dissipation is circularizing and removing them from the HB population. 
    \item While the cool star population shows evidence of tidal sculpting, the hot star population extends to, or near to, the wave-breaking limit, where tidal waves each periapse develop shocks and dissipate in turbulence near the stellar surface. This may be evidence of stellar evolution comparatively unchecked by tidal dissipation relative to the cool star population.  
    \item The lack of many systems at or above the wave breaking threshold indicates that this is a physical saturation point and that the lifetime of a wave-breaking phase is comparatively short. 
\end{enumerate}

In our analysis we have adopted many simplifications that allowed us to examine a large population of sources on the basis of their photometric data alone. On an individual basis, we caution that inferences for particular systems can be uncertain -- particularly in examples with other light curve features like eclipses. We suggest that our population-level analysis is an entry point to further study of particularly compelling cases from among this population by spectroscopic stellar and orbital characterization. The hot star population suffers from particularly high levels of parameter uncertainty from SED fitting alone since most of the photometry lies along the Rayleigh-Jeans tail of the spectrum. 

By the same token, we have made similar simplifications in our theoretical interpretation, for example, directly associating photometric and radial amplitudes, while ignoring possible coefficients \citep[e.g.][]{2022MNRAS.516.5021A} of order unity and any inclination effects. Some of these concerns can be addressed by future, more catered modeling of particular sources. We also anticipate that forward modeling the predicted HB population in the context of population synthesis models including tidal dissipation will be extremely useful in understanding the prevalence and amplitudes of HB stars as a function of their mass and evolutionary state, which relate to the nature of tidal dissipation.

\vspace{5mm}
This work would not be possible without the data collected by the OLGE team and made public by \citet{2022ApJS..259...16W,2022ApJ...928..135W}. We are deeply grateful for their contributions and commitment to making data both public and easy to work with. 
We are grateful to M. Kounkel for the SEDFit software and discussion of its use. We are grateful to A. Antoni, M. Cantiello, M. Gallegos-Garcia, B. Metzger, S. Millholland, and L. van Son  for feedback and helpful discussions.  
M.M. is supported by a Clay Postdoctoral Fellowship at the Smithsonian Astrophysical Observatory.

\software{  \texttt{astropy} \citep{astropy:2013, astropy:2018, astropy:2022}, \texttt{matplotlib} \citep{Hunter:2007}, \texttt{numpy} \citep{numpy}, \texttt{python} \citep{python}, \texttt{scipy} \citep{2020SciPy-NMeth, scipy_12522488}, \texttt{SEDFit} \citep{mkounkel_2023_10436982}, and the Software Citation Station \citep{software-citation-station-paper}.   }

\appendix 
\section{Reproduction Materials and Catalog}

The software and data to fully reproduce the figures and analysis in this work are available online at \url{https://github.com/morganemacleod/OGLE_HBs}. 

\begin{figure}
    \centering
    \includegraphics[width=0.45\linewidth]{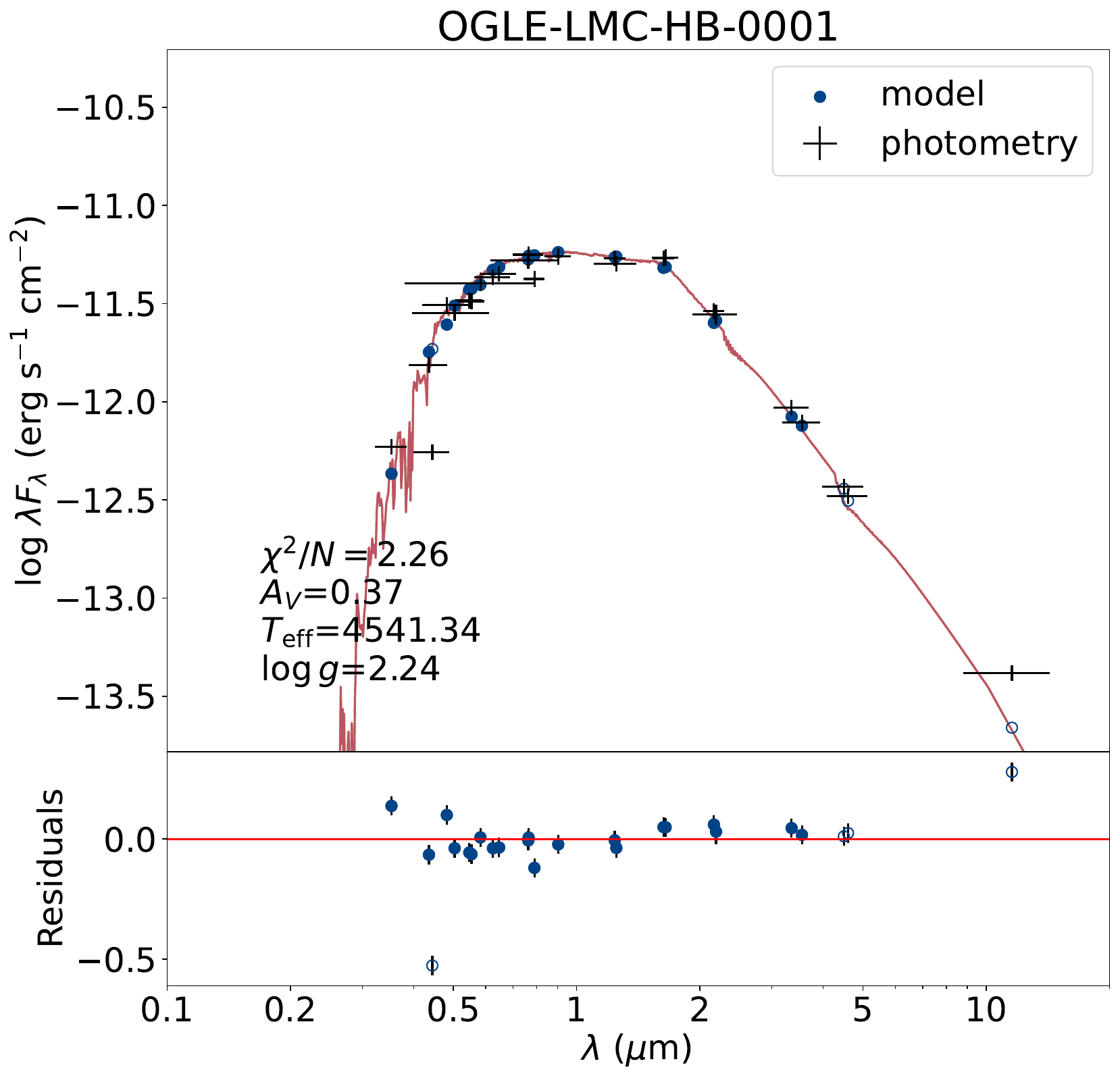}
    \includegraphics[width=0.45\linewidth]{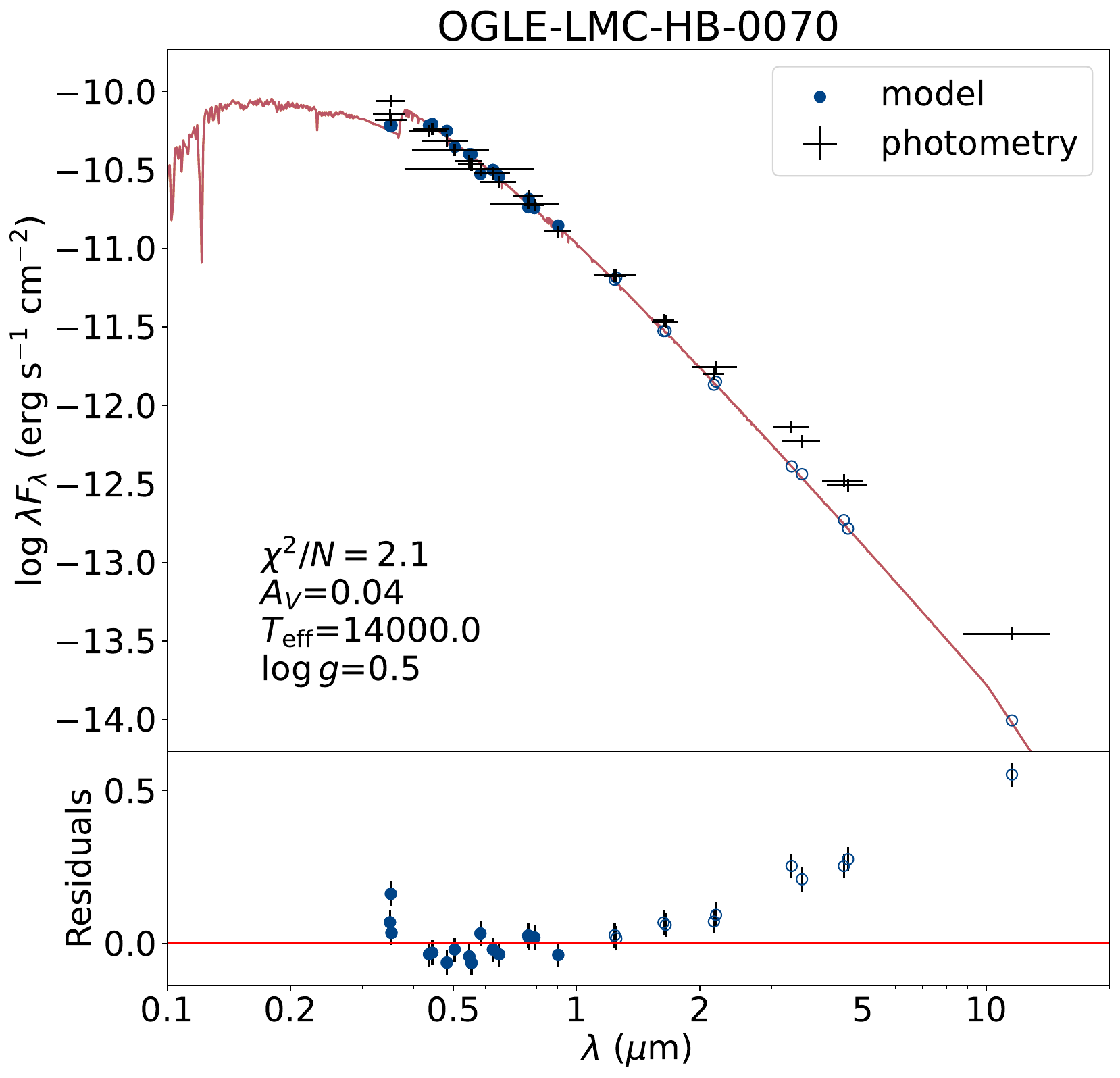}
    \caption{Example SED fits from analysis of broad-band photometry available on Vizier using {\tt SEDFit}. OGLE-LMC-HB-0001 is a cool star, with an SED that peaks in the optical/infrared range, while a hot star like OGLE-LMC-HB-0070 mostly shows data along the Rayleigh-Jeans tail. Solid points denote data that is fitted, unfilled points are data that is excluded from the fitting. In the case of OLGE-LMC-HB-0070, data above 1 micron are excluded because of the structured residual that shows evidence of infrared excess. }
    \label{fig:SEDFit}
\end{figure}

Examples of our SED fits based on Vizier photometry and \citet{mkounkel_2023_10436982}'s \texttt{SEDFit} software are shown in Figure \ref{fig:SEDFit}. One example shows a cool star with clearly resolved peak of the SED, while the other shows a hot star with photometry lying largely along the Rayleigh-Jeans tail. Both systems show excluded points as unfilled circles while fitted data are shown with filled points. The cool star shows an example of an outlier data point exclude in the subsequent analysis (we do not include data above 4 microns by default) and the hot star shows evidence for an infrared excess such that the fitting range was restricted to the optical by our algorithm as described in Section \ref{sec:method}.

We supplement the \citet{2022ApJS..259...16W,2022ApJ...928..135W} catalog of LMC and SMC sources, downloaded from the OGLE HB star query page at \url{https://ogledb.astrouw.edu.pl/~ogle/OCVS/hb_query.php}. This data is available as a machine readable {\tt .csv} file that can be most easily read with astropy's {\tt Table.read()} function. 
Each sources' light curve, including the \citet{2022ApJ...928..135W} fit of the \citet{1995ApJ...449..294K} model is available via the OGLE query page linked above.

\bibliographystyle{aasjournal}

\end{document}